\documentclass[aps,amssymb,amsmath,reprint]{revtex4-1} 

\usepackage{graphicx}  % needed for figures
\usepackage{dcolumn}   % needed for some tables
\usepackage{physics}
\usepackage{color}
\usepackage{appendix}
\usepackage{ulem}

\definecolor{orange}{rgb}{0.8, 0.3, 0}

\definecolor{blueviolet}{rgb}{0.2, 0.2, 0.6}
\usepackage[pdftex, % FOR PDFLATEX ONLY
bookmarks=true, % Bookmark bar
colorlinks=true,
allcolors=blueviolet,
pdfstartview={FitH}, % FitBH
]{hyperref}

\newcommand{\gens}{g_\text{ens}}
\newcommand{\bias}{I_\text{bias}}
\newcommand{\crystal}{\text{CaWO}_4}
\newcommand{\gaa}{\text{g}_{aa}}
\newcommand{\gbb}{\text{g}_{bb}}
\newcommand{\gcc}{\text{g}_{cc}}
\newcommand{\kc}{\kappa_\text{c}}
\newcommand{\ki}{\kappa_\text{i}}
\newcommand{\ktot}{\kappa}
\newcommand{\domega}{\Delta \omega}
\newcommand{\Tstar}{T_2^*}

\begin{document}

\title{Spin echo silencing using a current-biased frequency-tunable resonator}

\author{V.~Ranjan$^{1}$}
\email{vishal.ranjan@npl.co.uk}
\author{Y.~Wen$^{2}$}
\author{A.~K.~V.~Keyser$^{1,3}$}
\author{S.~E.~Kubatkin$^4$}
\author{A.~V.~Danilov$^4$}
\author{T.~Lindstr\"om$^{1}$}
\author{P.~Bertet$^{2}$}
\author{S.~E.~de~Graaf$^{1}$}
\email{sebastian.de.graaf@npl.co.uk}

\affiliation{$^1$National Physical Laboratory, Teddington TW11 0LW, United Kingdom }
\affiliation{$^{2}$Universit\'e Paris-Saclay, CEA, CNRS, SPEC, 91191 Gif-sur-Yvette Cedex, France} 
\affiliation{$^{3}$Imperial College London, Exhibition Road, SW7 2AZ, United Kingdom}
\affiliation{$^4$Department of Microtechnology and Nanoscience MC2, Chalmers University of Technology, SE-41296 Goteborg, Sweden}

%% Abstract #######################################
\begin{abstract} %% should be less than 600 characters incl spaces
The ability to control microwave emission from a spin ensemble is a requirement of several quantum memory protocols. Here, we demonstrate such ability by using a resonator whose frequency can be rapidly tuned with a bias current. We store excitations in an ensemble of rare-earth-ions and suppress on-demand the echo emission (`echo silencing') by two methods: 1) detuning the resonator during the spin rephasing, and 2) subjecting spins to magnetic field gradients generated by the bias current itself. We also show that spin coherence is preserved during silencing.
\end{abstract}

\maketitle
%%##################### Introduction #############################
Electron spins are a leading platform for implementing quantum memories both in the optical~\cite{lvovsky_optical_2009,hedges_efficient_2010,businger_optical_2020} and microwave~\cite{steger_quantum_2012,wolfowicz_atomic_2013,ortu_simultaneous_2018} domain, thanks to their long coherence times. Despite a relatively weak single spin-photon coupling, efficient absorption and emission of single photons~\cite{afzelius_proposal_2013,morton_storing_2018} can be reached if the spin concentration is high enough to reach the high-cooperativity regime ~\cite{imamoglu_cavity_2009,kubo_strong_2010, schuster_high-cooperativity_2010,abe_electron_2011,zhu_coherent_2011,amsuss_cavity_2011,ranjan_probing_2013,probst_anisotropic_2013,huebl_high_2013,sigillito_fast_2014,rose_coherent_2017,ball_loop-gap_2018}. Moreover, the inhomogeneous broadening of the spin ensemble provides numerous orthogonal degrees of freedom to allow multi-mode storage of quantum states using protocols based on the Hahn echo~\cite{wesenberg_quantum_2009,wu_storage_2010,grezes_multimode_2014,probst_microwave_2015,ranjan_multimode_2020}.   

An essential requirement of a quantum memory is random access, that is to retrieve a desired quantum state arbitrarily while keeping the others in the register intact until their subsequent retrieval [Fig.~\ref{fig:Qmemory}(a)]. The conventional two-pulse Hahn echo fails to fulfill this requirement since all stored states are simultaneously retrieved as last-in first-out. Moreover, the echo is emitted when the spins are all in the excited state, and thus it unavoidably gets superimposed with noise coming from spin spontaneous emission~\cite{julsgaard_quantum_2013,damon_revival_2011}. 

To achieve dynamic control of storage times and avoid population inversion during retrieval, various methods have been proposed and experimentally explored. Controlled and reversible inhomogeneous broadening (CRIB) implemented with electric or magnetic field gradients~\cite{kraus_quantum_2006,wu_storage_2010}, and cavity enhanced AC Stark shifts~\cite{zhong_nanophotonic_2017} can delay the emission of excitations on demand. Chirped control pulses for refocusing can also imprint phase gradients on the spin-ensemble to suppress the formation of an echo, and subsequently cancel the phase effecting a controlled retrieval~\cite{osullivan_random-access_2021, bonarota_photon_2014, minnegaliev_realization_2018}. 

\begin{figure}[t!]
  \includegraphics[width=\columnwidth]{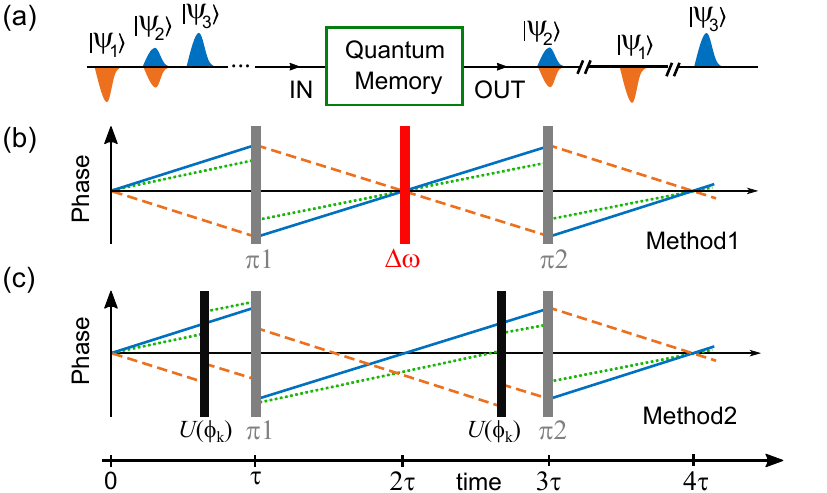}
  \caption{\label{fig:Qmemory}
  Quantum memory. (a) An ideal quantum memory architecture allowing retrieval of states at arbitrary times irrespective of the order during their storage. Blue and red reprsent two quadratures of the signal. (b) A sketch of phase evolution of spins (only three plotted for clarity) for a single excitation stored at time~=~0 to illustrate the proposed implementation of controlled suppression (at time $= 2\tau$) and retrieval (at $4\tau$) of echoes with fast detuning of the resonator frequency and (c) by generating an inhomogeneous phase encoding on spins.  The refocusing pulses $\pi 1$ and $\pi 2$ are ideal. Echo emission at $2\tau$ is not desirable since it contains noise from spontaneously emitting spins excited by $\theta 1$.}
\end{figure}

In this work, we demonstrate two methods of controlled suppression and retrieval of stored states by using current-biased fast frequency-tunable resonators [Fig.~\ref{fig:Qmemory}(b,c)]. The first method relies on the frequency tunability: after storage of an excitation and application of a refocusing pulse $\pi 1$, we rapidly detune the resonator during spin rephasing thus preventing the emission of an echo (referred throughout this work as `echo silencing'~\cite{julsgaard_fundamental_2013}). The detuning of the resonator is also useful for suppressing spin spontaneous emission by the Purcell effect~\cite{purcell_spontaneous_1946,bienfait_controlling_2016,eichler_electron_2017}, and thus for realizing a high-fidelity quantum memory. The second method uses the ability to pass a DC current through the resonator to implement a CRIB protocol. The current generates a magnetic field gradient imparting an inhomogeneous yet deterministic unitary phase evolution $U(\phi_k)$ on the $k^\text{th}$ spin, such that spins do not rephase and the echo is not formed [e.g. at time = 2$\tau$ in Fig.~\ref{fig:Qmemory}(c)]. In both methods, spin coherence is not affected by the act of echo silencing. Spins thus continue to precess until a second refocusing pulse $\pi 2$ (preceded by an identical phase evolution $U_k^\phi$ in Method2) triggers the rephasing of spins and emission of an echo. The echo retrieved at time~$=4\tau$ avoids noise as spins are now in the ground state~\cite{damon_revival_2011}.

%%######################### setup ##############################
Our measurements are made possible by superconducting resonators made of NbN whose kinetic inductance (inductor width 2$~\mu$m and thickness $50$~nm) allows for fast-tuning the resonance frequency $\omega_0$ when a DC current $\bias$ is passed through it (see Ref.~\cite{mahashabde_fast_2020} and supplementary~\cite{noauthor_notitle_nodate} for more details). The resonator performance is unaffected by the application of parallel magnetic fields up to 1~Tesla~\cite{graaf_magnetic_2012}, which is necessary to bring spins in resonance. Previous work by Asfaw \textit{et. al.}~\cite{asfaw_multi-frequency_2017} with frequency tunable resonators explored multi-frequency pulsed electron spin resonance experiments at a sample temperature of $T= 2$~K. In the following, we focus on aspects relevant to implementing a quantum memory, at 20~mK so that $\hbar \omega_0 \gg k_\text{B} T$.

The hybrid resonator-spin setup [Fig.~\ref{fig:hybridsetup}(a)] is inductively coupled to a feedline with a rate $\kc = 7.5 \times10^3 ~\text{s}^{-1} $, through which microwave signals are sent and received in transmission. The total cavity decay rate $\ktot = \kc+\ki$ is dominated by $\ki$, containing both dielectric losses and radiation losses through the current injection and exit terminals. We note that $\ki$ has input-power dependence due to the saturation of the two level system bath~\cite{mahashabde_fast_2020}. The change in the resonator frequency with $\bias$  [bottom right of Fig.~\ref{fig:hybridsetup}(a)] shows a quadratic response as expected from the kinetic inductance changes~\cite{vissers_frequency-tunable_2015,asfaw_multi-frequency_2017,mahashabde_fast_2020}. The crystal containing spins is glued on the resonator with vacuum grease. Two configurations are used: in ConfigI, the crystal is placed in a region far from the path taken by the DC current, whereas in ConfigII the crystal is directly above the path of current flow [red dashed lines in Fig.~\ref{fig:hybridsetup}(a)], and is thus maximally sensitive to magnetic field gradients.

\begin{figure}[t!]
  \includegraphics[width=\columnwidth]{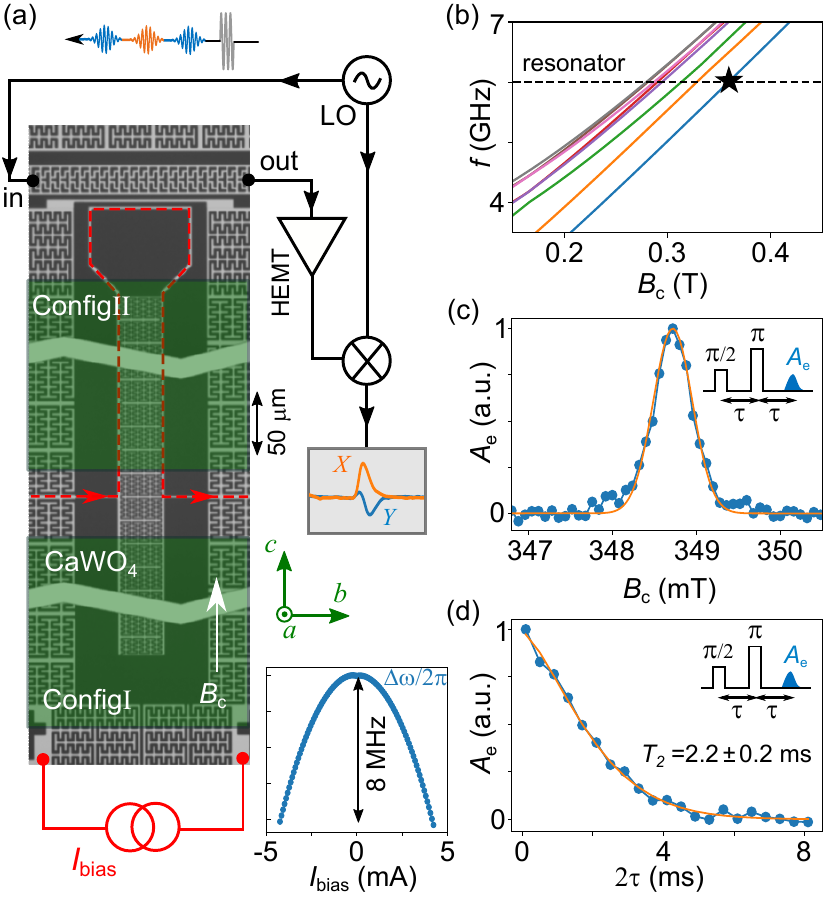}
  \caption{\label{fig:hybridsetup}
    Experimental setup. (a) An optical picture of the tunable resonator, the relative position of the $\crystal$ crystal, and a simplified measurement protocol. Horizontal zig-zag discontinuities in the picture hide the long repetitive vertical features. Signal quadratures $X,~Y$ are measured in transmission of the feedline to which the resonator is inductively coupled. The injection and exit routes of the bias current $\bias$ are shown by the dashed red line. Bottom right: the resonator frequency shift $\Delta \omega$ in response to $\bias$. (b) Numerically calculated electron spin transition frequencies for the $^{167}$Er isotope. (c) Echo-detected spin spectroscopy and (d) decoherence (measured: symbols, fits: lines) at the transition with $m_I = 7/2$ as marked by the star in Fig.~\ref{fig:hybridsetup}(b), where $m_I$ is the nuclear-spin projection on the $B$-field axis. }
\end{figure}

The spins are provided by an ensemble of Er$^{3+}$ ions in $\crystal$, with a nominal concentration 50~ppm ($6.4 \times 10^{17}~\text{cm}^{-3}$). Er$^{3+}$ substituting Ca$^{2+}$ ions in the lattice forms an effective electronic spin $S= 1/2$ system,  with an anisotropic \textbf{g}-tensor which is diagonal in the crystal frame ($\gaa = \gbb = 8.38$, $\gcc=1.25$). Moreover, roughly $23\%$ of dopants, namely the $^{167}$Er isotope possess a nuclear spin $I = 7/2$. Their spin Hamiltonian is $H_\text{Er-167}/\hbar =\mu_B \bold{S} \cdot \bold{g} \cdot \bold{B} - \bold{S} \cdot \bold{A} \cdot \bold{I} $, where $\mu_\text{B}$ is Bohr-magneton, $\bold{B}$ the applied magnetic field vector and \textbf{A} the hyperfine tensor that is also diagonal in the crystal frame ($A_{aa}/2\pi = A_{bb}/2\pi = 870~$MHz, $A_{cc}/2\pi =130~$MHz). Calculated electron spin transition frequencies for the $^{167}$Er isotope and a magnetic field aligned with the $c$-axis are plotted in Fig.~\ref{fig:hybridsetup}(b). 

Er$^{3+}$ spins are probed either in continuous wave (supplementary) or with a pulsed Hahn-echo sequence ($\theta/2-\tau-\theta - \tau-echo$). Since spins are located everywhere in the crystal, they undergo largely inhomogeneous Rabi rotation angles under the rectangular-shaped microwave pulses used throughout this work ($0.5~\mu$s duration).

%%##################### Echo characterization #############################
Echo-detected spectroscopy of the transition around 350~mT is plotted in Fig.~\ref{fig:hybridsetup}(c). Here and throughout, $A_e$ represents the integrated area of the echo. We find an approximate Gaussian lineshape with a full width half maximum (FWHM) of 0.6~mT or $\Gamma/2\pi=$10.5~MHz, which is much larger than the value due to dipolar couplings between spins for the nominal dopant concentration, $\sim 200$~kHz. This large broadening may arise from the inhomogeneous electric-field caused by charge-defects~\cite{mims_broadening_1966, dantec_twenty-threemillisecond_2021}. The echo magnitude $A_e$ as a function of $2\tau$ is shown in Fig.~\ref{fig:hybridsetup}(e); its decay is fit with a stretched exponential $\exp[-(2\tau/T_2)^x]$ yielding $T_2=2.2$~ms and $x=1.5$~\cite{rancic_electron-spin_2022}. Magnitude detection is employed to circumvent
the phase noise from the experimental setup. 

%%##################### Echo silencing #############################
We now utilize the resonator frequency tunability to demonstrate echo silencing. Echo traces for $\bias$ pulses, of varying amplitude (yielding different detuning $\domega$) and fixed $20~\mu$s duration applied around the time of echo emission [sketch in Fig.~\ref{fig:echo_silencing}(a)], are plotted in Fig.~\ref{fig:echo_silencing}(b). We observe a decrease of the echo magnitude with increased detuning. To analyze the spectral width of this decay, the echo area $A_e$ is plotted as as function of the detuning $\domega$ normalized to the resonator linewidth $\kappa$, where we have taken the high-power $\kappa/2\pi =0.14~$MHz. Numerical simulations assuming a uniform single spin-photon coupling strength $g_0$ are shown by solid line in Fig.~\ref{fig:echo_silencing}(c), and semi-quantitatively capture the decay. The discrepancy is due to limited bandwidth ($BW$) of the demodulation setup~\cite{noauthor_notitle_nodate-1}. The  simulated decay is well reproduced by the expression accounting for resonator filtering $\frac{\kappa/2}{\sqrt{\domega^2+\kappa^2/4} }$. The simulated echo-shapes are plotted in Fig.~\ref{fig:echo_silencing}(d), and show quantitative agreement with the experiment.

\begin{figure}[t!]
  \includegraphics[width=\columnwidth]{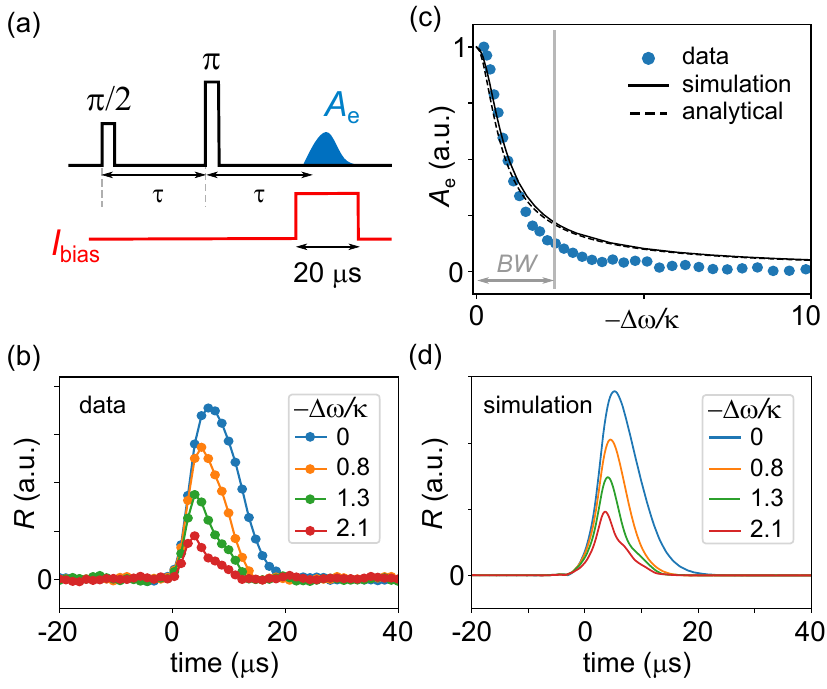}
  \caption{\label{fig:echo_silencing}
    Echo silencing with resonator tuning in configI. (a) Measurement scheme. Square $\bias$ pulses of fixed duration $20~\mu$s detune the resonator frequency by $\domega$ near the time of echo emission, $2\tau=200~\mu$s. (b) Echo shapes for various detuning. $R$ represents the signal magnitude $ \sqrt{X^2+Y^2}$. (c) Measured (symbols) and numerically simulated (line) Echo areas with increasing  detuning. Measurement bandwidth ($BW$) is also shown. (d) Numerically simulated echo shapes. }
\end{figure}

%%##################### Storage of weak pulses #############################
\begin{figure}[t!]
  \includegraphics[width=\columnwidth]{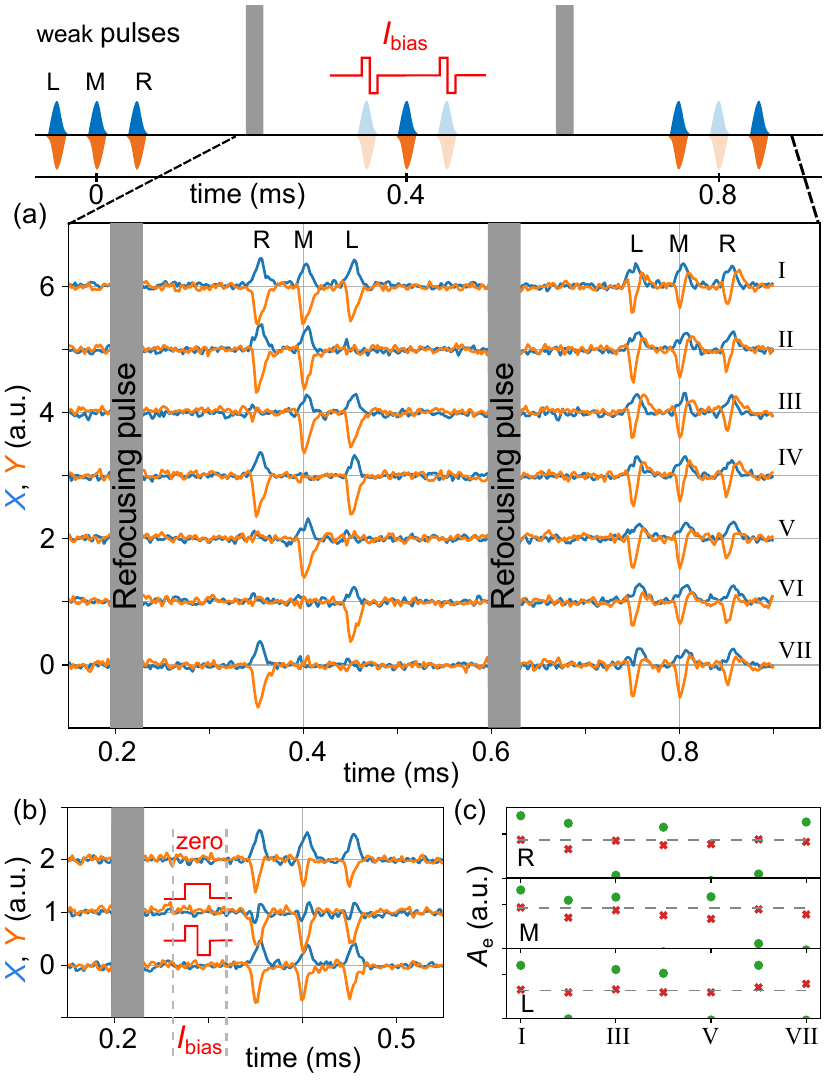}
  \caption{\label{fig:weakpulses}
    On-demand retrieval of weak pulses. (a) Signal quadratures of retrieval of three identical weak pulses for different permutations of echo suppression. The $\bias$ pulses of amplitude 2~mA and 20~$\mu$s total duration are made out of two halves of positive and negative current. Two refocusing pulses are applied along the same axis. (b) Comparison of retrieved quadratures for cases when no $\bias$ (top) or a $60~\mu$s long $\bias$ pulses centered at time=0.27~ms are applied with the same polarity (middle) or dual polarity (bottom) before the retrieval. (c) Magnitude of retrieved echoes, primary as circles and secondary as crosses, for cases numbered in panel (a). $\bias$ corresponds to a detuning of $15\kappa/2\pi=2~$MHz. Dashed horizontal lines represent signal magnitude of secondary echoes for the caseI with $\bias=0$. } 
\end{figure}

Having characterized the echo silencing, we now demonstrate coherent storage and retrieval of  microwave fields. Three identical Gaussian pulses $50~\mu$s apart and of FWHM = 4~$\mu$s (relative Rabi angle of $\pi/20$ and containing roughly $10^5$ photons) are sent and retrieved using the Hahn echo protocol [Fig.~\ref{fig:weakpulses}(a)]. Different permutations of echo suppression with square $\pm \bias$ pulses (of total duration $20~\mu$s, with two equal halves of positive and negative current yielding $\domega = -15\kappa$) are applied across primary echoes ($0.3~$ms $<$ time $< 0.5$~ms) to retrieve all, one or two echoes at a time. We find that retrieved echo magnitudes and phases are not affected by the echo suppression preceding them. Minor discrepancies can be attributed to the phase noise from the setup. 

The $\bias$ pulses in Fig.~\ref{fig:weakpulses}(a) are made of two halves of positive and negative current to cancel the associated inhomogeneous magnetic field gradients (see supplementary). To show its importance, $60~\mu$s long $\bias$ pulses are applied in-between the refocusing pulse and echoes. As shown in Fig.~\ref{fig:weakpulses}(b), a $\bias$ pulse of single polarity induces a visible phase-shift in the echo while the original echo phase is recovered when applying dual polarity.

To quantify the memory performance, the field retrieval efficiency of $7 \times 10^{-3}$ is obtained by comparing the integrated input fields with echoes (supplementary). This value is somewhat lower than the theoretical upper bound~\cite{afzelius_proposal_2013} $\frac{4C}{(1+C)(1-C)}(\kc/\ktot) = 1.1 \times 10^{-2}$ likely due to an inefficient refocusing pulse caused by spatially inhomogeneous Rabi angles. Here the cooperativity $C=4\gens^2/\ktot \Gamma = 0.23$ is deduced from continuous wave transmission spectroscopy yielding the ensemble coupling strength $\gens/2\pi = 0.35~$MHz. The low efficiency observed here, despite $C=0.23$, is due to a relatively large radiation loss through bias-current injection/exit points yielding a ratio $\kc/\kappa \approx 100$~\cite{noauthor_notitle_nodate}.

We also apply an identical second refocusing pulse and retrieve secondary echoes (time $> 0.7$~ms) in different cases of echo silencing. For comparison, we have plotted the echo magnitudes in Fig.~\ref{fig:weakpulses}(c).  No visible changes, neither in the magnitude or phase, are observed in secondary echoes when corresponding primary echoes are suppressed. This is again due to the low efficiency of our memory protocol. 

%############## CRIB #########################
\begin{figure}[t!]
  \includegraphics[width=\columnwidth]{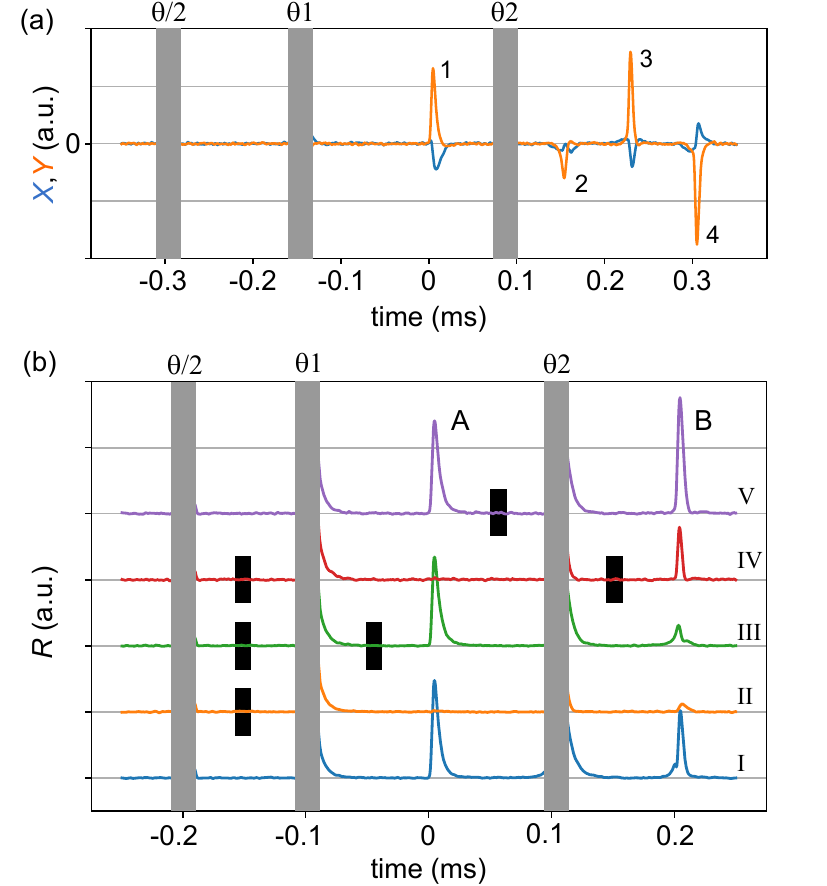}
  \caption {\label{fig:CRIB}
    Echo silencing with magnetic field gradients in configII.  (a) Signal quadratures of a train of echoes created by unevenly spaced control pulses and $\bias=0$. echo1 (echo2) is primary (secondary) rephasing of spin coherence created by $\theta/2$, echo3 is 3-pulse stimulated echo, and echo4 is primary rephasing of spin coherence created by $\theta 1$. (b) Signal magnitude of two echoes measured with different permutations of $\bias$ pulses. Each bias pulse is of single polarity, duration $20~\mu$s and amplitude 3~mA (positions shown as black rectangles). The curves have been vertically offset for clarity. In both panels, all control pulses are applied along the $x$-axis, with Rabi angles of $\theta = \theta 1 = \theta 2 $. }
\end{figure}

The inhomogeneous Rabi angles in our setup also lead to stimulated echoes (SE), affecting state retrieval after multiple control pulses. An example of SE can be identified as echo3 in the measurement shown in Fig.~\ref{fig:CRIB}(a), whereas echo1 and echo4 are primary Hahn echoes, and echo2 a secondary Hahn echo. In the following, we show that using a suitable combination of magnetic field gradients~\cite{wu_storage_2010} from $\bias$ (measurements acquired in configII), we can selectively suppress primary echo, and/or secondary echo, and/or stimulated echo. 

Echo signal magnitude after two cycles of refocusing  and different permutations of $\bias$ pulses is shown in Fig.~\ref{fig:CRIB}(b). Due to a delay $2\tau$ between $\theta 1$ and $\theta 2$, contributions from secondary echo and SE are superposed in echoB. As a reference, caseI is measured with $\bias=0$. In caseII, a single bias pulse randomizes the coherence created by the $\theta/2$ pulse and therefore both echoes are suppressed. By applying another identical $\bias$ pulse after $\theta 1$, echoA is revived (caseIII) but echoB is only partially recovered. The latter is due to the dominant contribution from SE which is absent in caseII and caseIII. SE  is however revived when another $\bias$ pulse is applied after $\theta 2$ as shown in caseIV. Note that phase evolution of spins producing SE can be constructed by replacing $\theta 1-2\tau-\theta 2$ by an effective refocusing pulse in the sketch shown in Fig.~\ref{fig:Qmemory}(c). In caseV, echoB is even larger compared to caseI due to suppression of the secondary Hahn echo which would otherwise have a sign opposite to that of the SE~[Fig.~\ref{fig:CRIB}(a)]. Overall, these measurements demonstrate the use of a local on-chip source of magnetic field gradients to implement CRIB~\cite{wu_storage_2010}. 

To quantify the performance of our CRIB protocol, we compare the area under echoes in Fig.~\ref{fig:CRIB}(b). For echoA, we find the suppression (revival) fidelity to be $98\% ~(96\%)$. For echoB, suppression fidelity is $84\%$, while retrieval fidelity is $62\%$ and $33\%$ for secondary and stimulated echoes, respectively. The reduced fidelity for echoB is due to growing pulse error and highlights the importance for efficient refocusing needed in an ideal memory protocol [Fig.~\ref{fig:Qmemory}(c)]. 

In summary, we have demonstrated the use of a current biased tunable resonator for echo suppression, phase preserving retrieval of states and controlled reversible inhomogeneous broadening in an ensemble of Er spins. Future efforts towards implementing the quantum memory will aim at unit efficiency, explore optimized resonator designs to mitigate radiation losses from crystal mounting, and adopt a spatially localized spin ensemble to attain efficient refocusing from uniform Rabi angles~\cite{ranjan_spatially_2021}.

%% ############ Acknowledgement
We acknowledge the support of the UK government department for Business, Energy and Industrial Strategy through the UK national quantum technologies program. AD and SK acknowledge the support from the Swedish Research Council (VR) (grant agreements 2020-04393 and 2019-05480).

%##########  Bibliography  ######################
%\bibliography{QuantumMemory}


%\documentclass[aps,amssymb,amsmath,reprint]{revtex4-1} 
%\usepackage{graphicx}  % needed for figures
%\usepackage{dcolumn}   % needed for some tables
%\usepackage{physics}
%\usepackage{color}
%\usepackage{appendix}
%\usepackage{ulem}

%\definecolor{blueviolet}{rgb}{0.2, 0.2, 0.6}
%\usepackage[pdftex, % FOR PDFLATEX ONLY
%bookmarks=true, % Bookmark bar
%colorlinks=true,
%allcolors=blueviolet,
%pdfstartview={FitH}, % FitBH
%]{hyperref}

%\newcommand{\gens}{g_\text{ens}}
%\newcommand{\bias}{I_\text{bias}}
%\newcommand{\crystal}{\text{CaWO}_4}
%\newcommand{\gaa}{\text{g}_{aa}}
%\newcommand{\gbb}{\text{g}_{bb}}
%\newcommand{\gcc}{\text{g}_{cc}}
%\newcommand{\kc}{\kappa_\text{c}}
%\newcommand{\ki}{\kappa_\text{i}}
%\newcommand{\ktot}{\kappa}
%\newcommand{\Tstar}{T_2^*}

\newcommand{\pbcor}[1]{{\color[rgb]{0.0,0.5,0.0}#1}} % Patrice's edits in Green

%################### SUPPLEMENTARY ###################
\renewcommand{\theequation}{S\arabic{equation}}
\renewcommand{\thefigure}{S\arabic{figure}}
\renewcommand{\thetable}{S\arabic{table}}

\setcounter{equation}{0}
\setcounter{figure}{0}

%\begin{document}

\title{Supplementary material to ``Echo silencing using a current biased tunable resonator"}

%\author{V.~Ranjan$^{1}$}
%\email{vishal.ranjan@npl.co.uk}
%\author{Y.~Wen$^{2}$}
%\author{A.~Keyser$^{1}$}
%\author{S.~Kubatkin$^3$}
%\author{A.~Danilov$^3$}
%\author{T. Lindstr\"om$^{1}$}
%\author{P.~Bertet$^{2}$}
%\author{S. E. de Graaf$^{1}$}
%\email{sebastian.de.graaf@npl.co.uk}

%\affiliation{$^1$National Physical Laboratory, Teddington TW11 0LW, United Kingdom }
%\affiliation{$^{2}$Universit\'e Paris-Saclay, CEA, CNRS, SPEC, 91191 Gif-sur-Yvette Cedex, France} 
%\affiliation{$^3$ Department of Microtechnology and Nanoscience MC2, Chalmers University of Technology, SE-41296 Goteborg, Sweden}

\maketitle

%%######################### setup ##############################
\section{Measurement setup}
The measurements are performed at the base temperature of the dilution refrigerator at 20~mK, unless stated otherwise. A high electron mobility transistor (HEMT) amplifier mounted at 4~K stage  forms the first amplifier in the detection chain. Signals are further amplified at room temperature both before and after the demodulation to an intermediate frequency of 40~MHz with a 6~dB demodulation bandwidth of $\sim$300~kHz (Zurich Instruments Lock-in amplifier).

The bias current $\bias$ is injected into the resonator through the split ground planes. On the injection side, thermalization and filtering are provided by a 20~dB attenuator at the 4~K stage and low pass filters (home made ecosorb and commercial mini-circuits filters installed in series with a 3~dB roll off at 100~MHz) at the 20~mK stage. The current returns to a 50~ohm termination at the 4~K stage. The analog output of a Tektronix arbitrary waveform generator is used for fast tuning of $\bias$.

%%######################### Resonator ##############################
\section{Resonator properties}
\begin{figure}[t!]
  \includegraphics[width=\columnwidth]{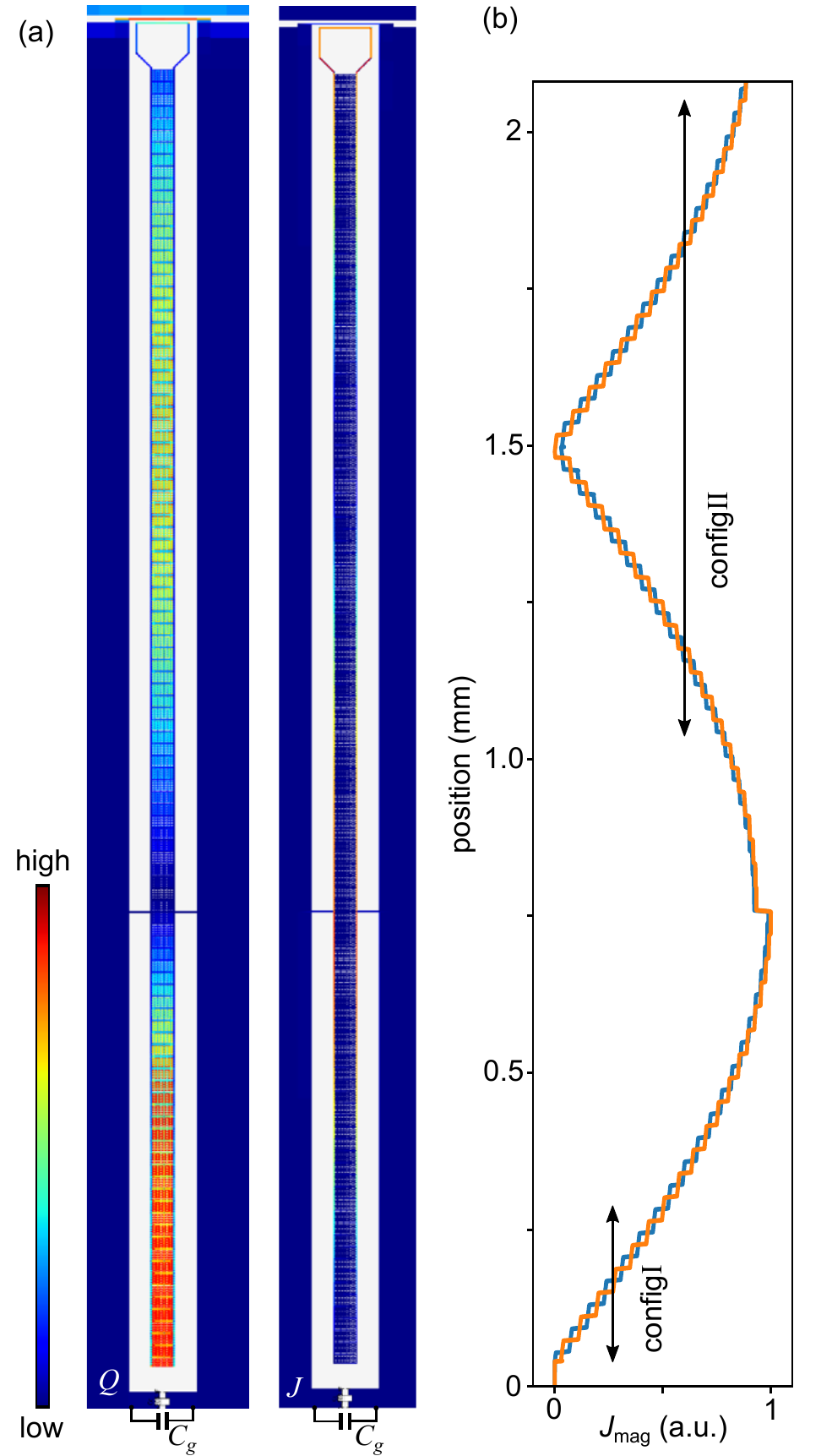}
  \caption{\label{Sfig:sonnet} Sonnet simulation of the resonator showing (a) charge density ($Q$) and current-density ($J$) distribution. (b) Current density through two vertical prongs of the resonator. The kink in $J_\text{mag}$ at the $\bias$ injection highlights the radiation loss. Crystal coverage for two configurations used in this work is illustrated by double arrows.
}
\end{figure}

The design and operation of our resonators are described in details in Ref.~\cite{mahashabde_fast_2020}. In the following, we summarize the main points. The resonator design, depicted in Fig.~\ref{Sfig:sonnet}, is an electromagnetic analog of mechanical tuning fork: U-shaped superconducting line, 2~$\mu$m wide, operates as an inductor; the turning point inductively couples to the feedline. The area in-between the prongs is filled with fractalized prong-to-prong capacitor, patterned with 1~$\mu$m design rule. We exploit the $3\lambda/4$ mode, which supports voltage node (current antinode) at a distance $\lambda/4$ from the open end, as shown by sonnet simulations in Fig.~\ref{Sfig:sonnet}. The bias current, used to control the kinetic inductance, is injected via control terminals which couple at the voltage node points. To allow for DC bias, the ground plane is split into two half-planes which couple through the fractalized capacitor $C_g$. In Sonnet simulation, a simplified model was used: the fractalized capacitor was substituted with lumped element capacitor of value $C_g=4$~pF, and the ground elements, also fractalized in a real design, were replaced with solid polygons. We find that a kinetic inductance contribution of $3.19~$pH/$\square$ is needed to account for the measured bare resonator frequency of 6.13~GHz. We furthermore extract the characteristic impedance of $87~\Omega$, from which we estimate the single spin-photon coupling strength $g_0/2\pi=25~$Hz at the current antinode and a height above the resonator plane of $2~\mu$m.

An important implication of mounting the $\crystal$ crystal (dielectric constant of $\sim 10$) on these resonators is redistribution of the electromagnetic mode. This may lead to two effects. Firstly, radiation losses through current injection and exit points can become appreciable as shown in the current density ($J$) plot of Fig.~\ref{Sfig:sonnet}(b). Secondly, the coupling to the feedline can also change, yielding different $\kappa_c$. The exact simulation of resonator-crystal setup could not be done due to demanding fractal architecture of the ground plane.

\begin{figure}[t!]
  \includegraphics[width=\columnwidth]{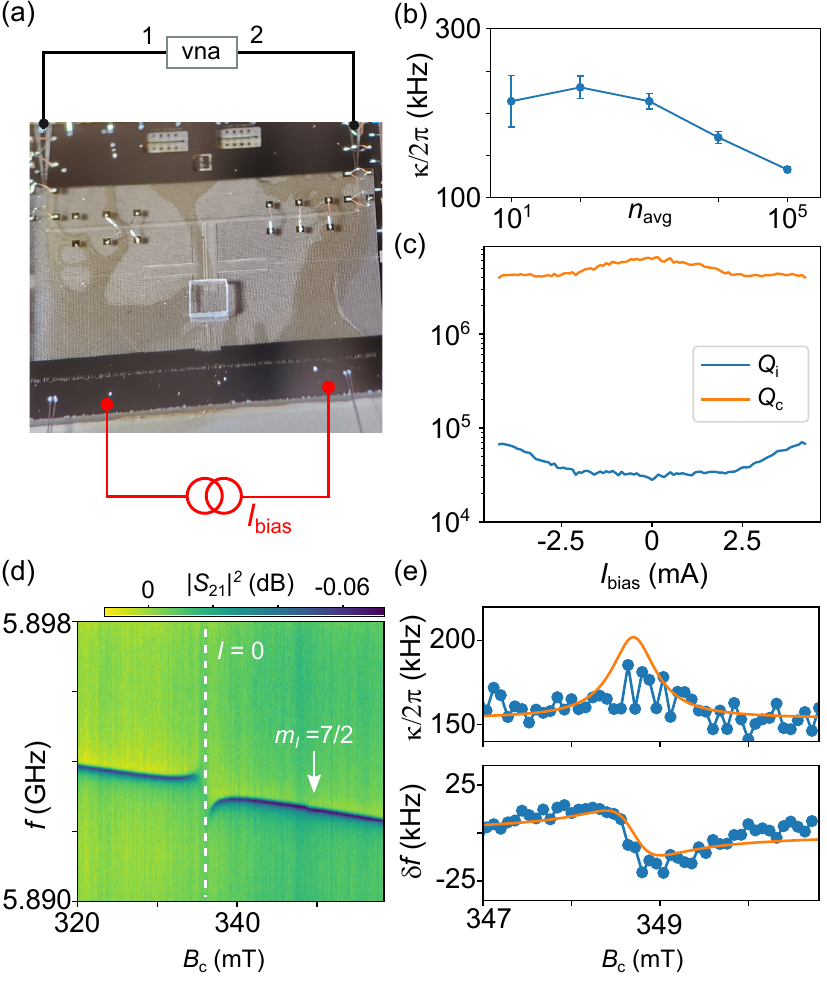}
  \caption{\label{Sfig:hybridsetup}
   VNA spectroscopy with crystal configuration configI.  (a) An optical image of the resonator with the crystal mounted near the open-end of the resonator. Signals are acquired in transmission $S_\text{21}$ of the feedline. (b) Total loss rate $\kappa$ with average intra-cavity photon number. (c) Extracted quality factors versus bias current $\bias$. (d) VNA spectroscopy at an average intra-cavity photon number of $10^4$. (e) Extracted resonator linewidth $\kappa$  and frequency shift $\delta f$ versus $B_c$. Solid lines are numerically calculated from Eq.~\ref{eq:s21}. We have subtracted a linear background of frequency shift due to the kinetic inductance of NbN. Measurements in panel(b,c) are done at 348.7~mT.}
\end{figure}

The $\crystal$ crystal has a surface area of 1~mm$^2$ and thickness $0.2~$mm and held on top of the resonator with the help of vacuum grease. The crystal is positioned in the lower half or upper half of the resonator, namely configI and configII. For configI (configII), bare resonator frequency shifted by 220~(80)~MHz. The extrinsic coupling rate to the feedline $\kappa_c$ reduced by a factor of 3 compared to the case of no crystal for configI, while it was larger by a factor of 5 for configII. 

We note that our fractal resonators in general couple strongly to a bath of two level systems (TLSs) residing on the substrate. With increasing powers TLSs get saturated and hence $\ki$ becomes dependent on number of intra-cavity photons. In our case, this change in $\kappa_i/2\pi$ 0.24~kHz and 0.14~kHz for $10^1-10^5$ intra-cavity photons is relatively weak compared to previously measured in Ref.~\onlinecite{mahashabde_fast_2020}. Moreover, measured internal $Q_i$ and external $Q_c$ quality factors show an increase with $\bias$ [Fig.~\ref{Sfig:hybridsetup}(c)]. These observations hint at a possible role of radiation losses.

%%######################### van spectroscopy ##############################
\section{Extraction of $\gens$}
We perform continuous wave spectroscopy, done using a vector network analyzer (VNA), to extract the spin ensemble coupling strength $\gens$. The complex transmission through the feedline can be derived using input output theory and reads~\cite{schuster_high-cooperativity_2010}

\begin{equation} \label{eq:s21}
    S_\text{21} = 1- \frac{\kc}{\kappa + 2i\Delta_0+ \frac{4\gens^2}{2i\Delta_s +\Gamma}},
\end{equation}
where $\Gamma$ is the spin linewidth, and  $\Delta_s$ ($\Delta_0$) is the detuning of spin transition (resonator) frequency from the probe frequency.

In the VNA spectroscopy shown in Fig.~\ref{Sfig:hybridsetup}(d), no noticeable absorption is observed near $m_I=7/2$. The analysis is complicated by the background and shallow depth of the resonance.  We can however use the frequency shift data to quantify $\gens$. We note that the small changes in $\ki$ actually allow for a reliable extraction of the spin linewidth $\Gamma/2\pi = 10.5~$MHz from echo measurements (Fig.~2(c) of the main text). The solid line with $\gens/2\pi=0.35~$MHz calculated with Eq.~\ref{eq:s21} describes the measured data in Fig.~\ref{Sfig:hybridsetup}(d) well. 

\begin{figure}[t!]
  \includegraphics[width=\columnwidth]{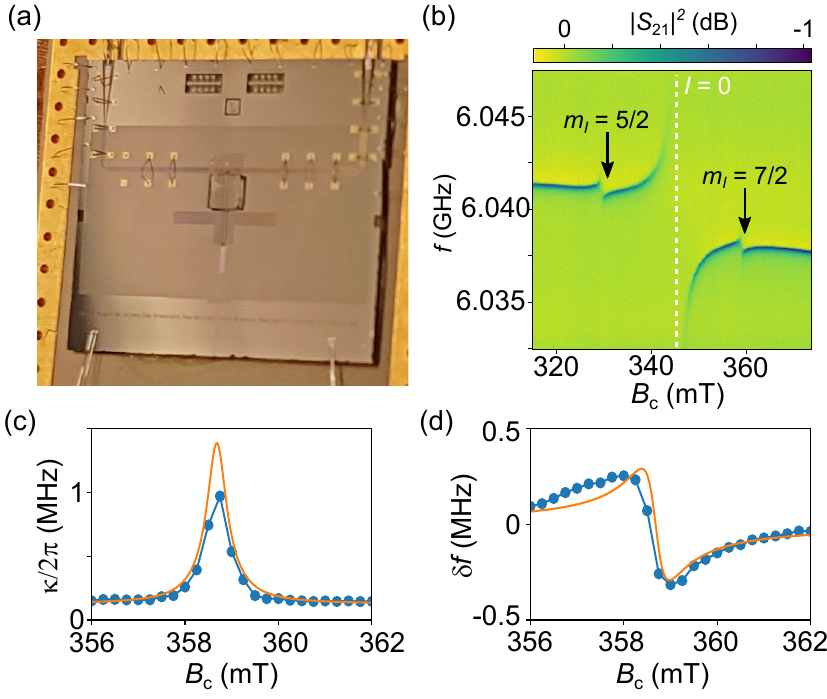}
  \caption{\label{Sfig:hybridsetup2}
   vna spectroscopy with crystal configuration configII.  (a) An optical image of the resonator with crystal mounted on the part of the resonator with maximum RF current. (b) Continuous wave spectroscopy done with VNA at an average intra-cavity photon number of $10^4$. (c) Extracted resonator linewidth $\kappa$  and (d) and frequency shift $\delta f$ versus $B_c$. Solid lines in panel(c,d) are numerically calculated from Eq.~\ref{eq:s21}. }
\end{figure}

We follow a similar analysis for the crystal configuration configII (Fig.~\ref{Sfig:hybridsetup2}). The VNA spectroscopy now shows clear absorption and dispersion near the same spin transition with $m_I=7/2$. The larger value of extracted $\gens/2\pi=1.8~$MHz can be explained by stronger $B_1$ field in the top of the resonator on average and better crystal coverage. Note that we expect the electron spin transition with $m_I=7/2$ to occur at 361.4~mT for the magnetic field strictly aligned with crystal $c$-axis suggesting a misalignment of 3.3~degrees, which could arise from errors in crystal axis during cutting, crystal positioning and due to misalignment of the applied field.

%%%%%%%%%%%%%%% Coherence times %%%%%%%%%%%%%%%%%%%%%%%%
\section{$T_1$ and $T_2$ times}
\begin{figure}[t!]
  \includegraphics[width=\columnwidth]{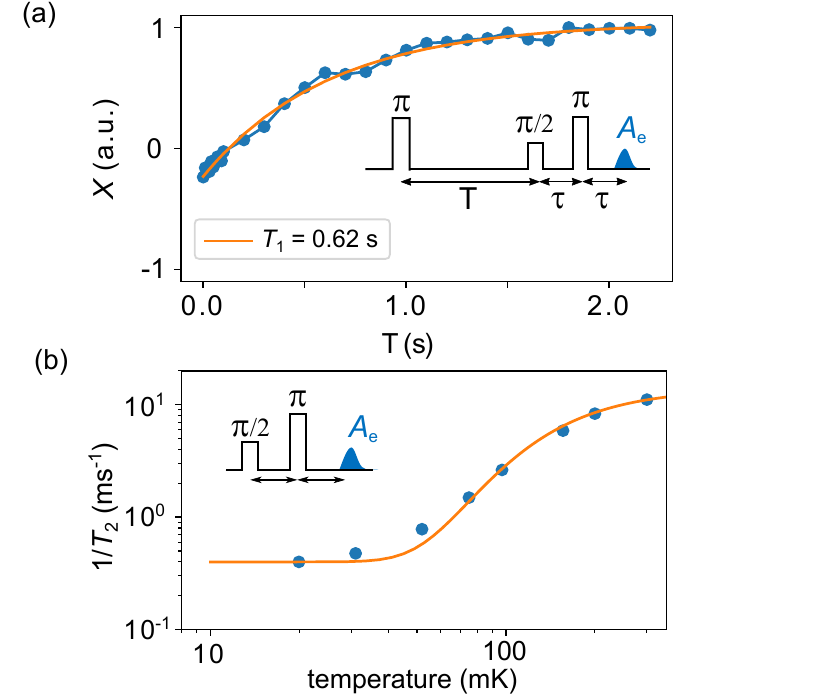}
  \caption{\label{Sfig:T2}
    Spin decoherence and relaxation at the spin transition $m_I=7/2$. (a) Inversion recovery spin-relaxation measurement (data: symbols and fit: line) yielding $T_1 = 0.62 \pm 0.15$~s. (b) Symbols: Temperature dependence of inverse of coherence times measured using Hahn echo sequence. Solid line: Calculation using Eq.~\ref{eq:SD}.  }
\end{figure}

We use an inversion recovery sequence to obtain the spin energy relaxation time of $T_1 = 0.62 \pm 0.15$~s [Fig.~\ref{Sfig:T2}(a)]. It is interesting to compare this with the value measured in Ref.~\cite{marianne_le_dantec_electron_2022} where the applied magnetic field was also parallel to the c-axis: 0.2~s at a frequency $\omega_0 / 2\pi = 7.85~$GHz. For the direct phonon process, $T_1$ should scale as $\omega_0^{-5}$~\cite{larson_spin-lattice_1966}, and our measurements are indeed compatible with this scaling, indicating direct phonon as the main relaxation mechanism.

We observed $T_1$ to decrease with lower power of $\pi$ pulses, reaching values as low as $0.12~$s. Indeed such effects are expected in the Purcell regime because of increasing $g_0$ with decreasing distance from the resonator~[see  Ref.~\onlinecite{ranjan_pulsed_2020,dantec_twenty-threemillisecond_2021}]. However, the expected Purcell relaxation time $\kappa/4g_0^2$ in our case is $20-80$~s, for coupling strengths ranging between $g_0/2\pi = 10-20$~Hz. We believe that spin-diffusion resulting in spins escaping the detection-volume of the resonator could be a likely process determining $T_1$ of our measurements at lower powers. This escape should be dependent on the total volume of spins probed which decreases with increasing $g_0$ or decreasing power. 

We performed spin coherence time $T_2$ measurements for the $m_I= 7/2$ transition at various temperatures [see Fig.~\ref{Sfig:T2}(b)], and find a strong dependence. A likely process affecting coherence times in heavily-doped crystals is spectral diffusion caused by flip-flop between dopant spins. In our case, Er-isotope with zero nuclear spin ($\text{Er}^0$) forms the largest paramagnetic centers (visible in the continuous wave spectroscopy as a large avoided crossing). The associated spin-flips versus temperature $T$ are known to follow Boltzmann statistics such that $T_2$ of central spins scales as~\cite{bai_time_1989}

\begin{equation} \label{eq:SD}
    1/T_2 = 1/T_2^0 + R_S \sech^2 \left( \frac{\text{g} \mu_\text{B} B_c}{2 k_\text{B} T}\right),
\end{equation}
where $\mu_\text{B}$ is the Bohr magneton, $k_\text{B}$ the Boltzmann constant, $R_s$ the maximum rate contribution from flip-flops being dependent on the dopant-concentration, and $1/T_2^0$ the residual decoherence rate from other factors. Calculation using Eq.~\ref{eq:SD} is shown as solid lines in Fig.~\ref{Sfig:T2}(b). We find a good match with the data over the entire temperature range. Similar observations were made in a recent work done on a sample with comparable doping~\cite{rancic_electron-spin_2022}. 

The residual coherence time $T_2^0 = 2.5~$ms measured at the lowest temperature is an order of magnitude smaller than 23~ms measured in a natural abundance crystal, where it was limited by the spectral diffusion due to the nuclear spin bath~\cite{marianne_le_dantec_electron_2022}. We think instantaneous diffusion might be contributing to the reduced coherence, though it could not be ascertained due to poor refocusing pulses and power dependence of the resonator bandwidth.

%%%%%%%%%%%%%%%%%% Theory %%%%%%%%%%%%%%%%%%%%%%%%%%
\section{Theory and simulations}
For numerical simulations presented in this work, we employ input-output theory in the rotating frame to model the signal emitted from $N$ spins coupled to a resonator. The intra-cavity field $\alpha$ is given by

\begin{equation}
    \frac{d\alpha}{dt} = \sqrt{\kappa_c}\beta - (\kappa/2 + i\Delta_0)\alpha - i\sum^N_{m=1} g_0 \langle \hat{\sigma}_m^{+} \rangle,
\end{equation}
where the index $m$ represents a discrete spin, $\beta$ the drive amplitude, $\Delta_0$ the detuning of drive frequency from the resonator, $\kappa_c$ and $\kappa$ being the coupling and total decay rate, respectively. The transverse component of spins, given by the expectation value of the operator $\hat{\sigma}^\pm = (\hat{\sigma}_x \pm \hat{\sigma}_y)/2$, populates the resonator at a rate $g_0$, which we assume to be same for all spins. Here, $\hat{\sigma}_x$ and $\hat{\sigma}_y$ are standard Pauli matrix operators. The time evolution of each spin's density matrix $\rho$ is described by the Lindblad equation as

\begin{equation}
    \frac{d\rho}{dt}=-\frac{i}{\hbar}[\hat{H}_0,\rho] +  \hat{\mathcal L}_\text{1}(\rho) +  \hat{\mathcal L}_\text{2}(\rho), 
\end{equation}
where the Liouvillian super-operators $\hat{\mathcal L}_1, ~\hat{\mathcal L}_2$  account for spin-relaxation and decoherence, respectively, and are given by $\hat{\mathcal L}_i(\rho) = \mathcal L_i \rho \mathcal L_i^{\dagger} - \{\mathcal L_i^{\dagger} \mathcal L_i, \rho \} $ with $\mathcal L_1 = \sqrt{1/T_1}\hat{\sigma}^{-}$ and $\mathcal L_2 = \sqrt{1/2T_2}\hat{\sigma}_z$. The Hamiltonian of each spin is described by 

\begin{equation}
H_0/\hbar = \frac{\Delta_s}{2}\hat{\sigma}_z + g_0(\alpha \hat{\sigma}^{+} + \alpha^* \hat{\sigma}^{-} ),     
\end{equation}
where $\Delta_s$ is the detuning of spin's Larmor frequency from the resonator.

%%%%%%%%%%%%%%% Echo Silencing %%%%%%%%%%%%%%%%%%%%%
\section{Echo silencing}
\begin{figure}[t!]
  \includegraphics[width=\columnwidth]{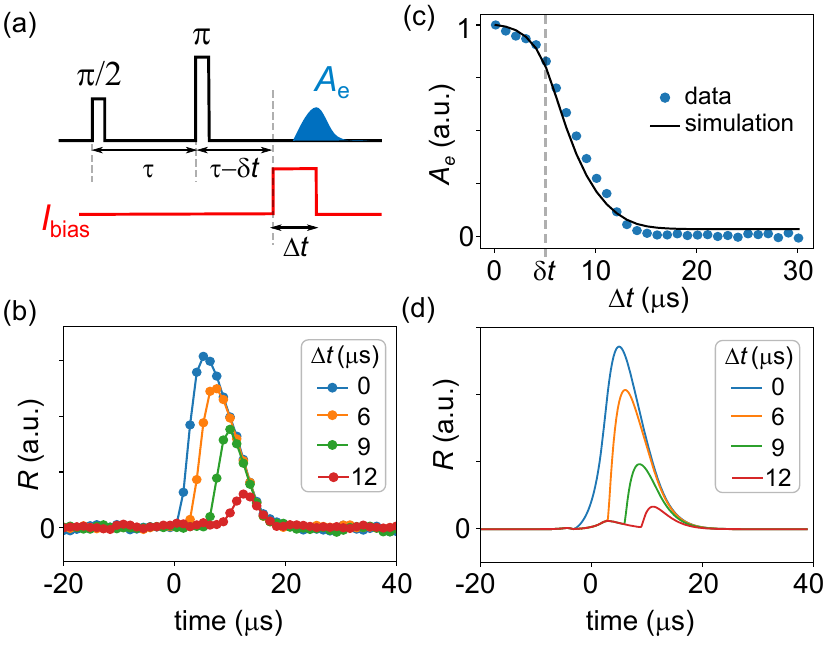}
  \caption{\label{Sfig:echo_silencing}
    Partial suppression of echo in configI. (a) Measurement scheme. Square $\bias$ pulses corresponding to a detuning of $-15\kappa$, offset by fixed $\delta t=5~\mu$s from the time of echo ($2\tau$) and of varying duration $\Delta t$ is applied around the echo. (b) Family of echoes measured at different $\Delta t$. $R$ represents the signal magnitude $ \sqrt{X^2+Y^2}$. (c) Measured (symbols) and simulated (line) echo area decay versus $\Delta t$. (d) Numerical simulation of echo shapes for varying $\Delta t$. In panel (b,d) time axis is relative to $2\tau$. }
\end{figure}

In addition to the measurement presented in the main text (Fig.~3), we also studied the effect of suppressing part of the echo on the final echo shape and magnitude. To this end, square $\bias$ pulses corresponding to a detuning of 2~MHz ($\approx 15 \kappa/2\pi$) and varying duration $\Delta t$ are applied near the echo. As sketched in Fig~\ref{Sfig:echo_silencing}(a), the bias pulse is already turned on at a time $\delta t = 5~\mu s$ before $2\tau$. A family of echo traces for different $t_0$ is plotted in Fig.~\ref{Sfig:echo_silencing}(b) and shows a continual decrease of echo magnitudes with increasing $\Delta t$. The integrated echo signal $A_e$  [Fig.~\ref{Sfig:echo_silencing}(b)] furthermore shows a complete suppression when the end of the bias pulse reaches half the size of the echo ($\Delta t-\delta t \approx 10~\mu$s). Numerical simulations shown as solid lines [Fig.~\ref{Sfig:echo_silencing}(c,d)] show quantitative agreement with the decay and echo shapes. Slight mismatch ($< 4\%$) between data and simulation at large $\Delta t$ is due to finite measurement bandwidth of our setup.

%############## Ibias dephasing #########################
\section{Dephasing from $\bias$} \label{sec:dephasing}

\begin{figure*}[t!]
  \includegraphics[width=2\columnwidth]{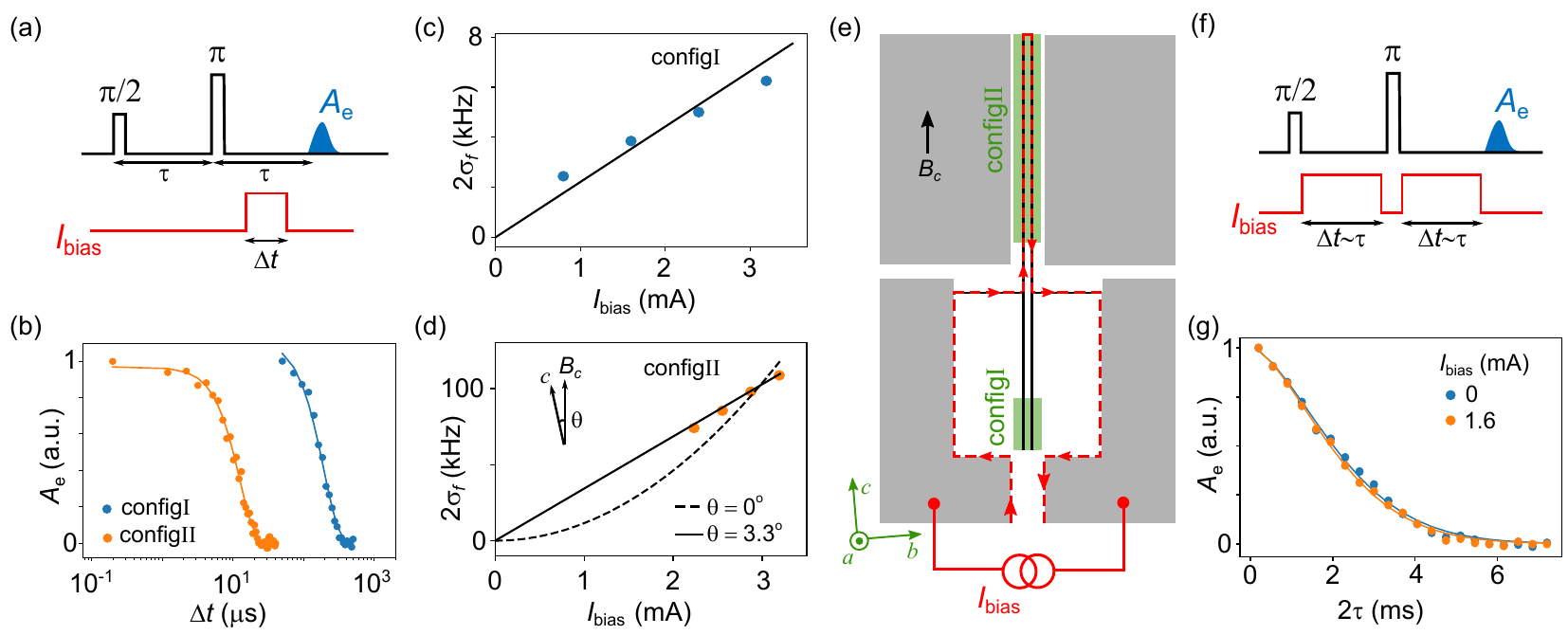}
  \caption {\label{Sfig:dephasing}
    Dephasing from $\bias$. (a) Measurement scheme to deduce dephasing from $\bias$. The duration of bias pulse $\Delta t \lg \tau$ (b) Echo magnitude decay (measurement: symbols,  fits:lines) for two configurations measured with the same $\bias=2.4~$mA. (c) Inhomogeneous broadening ($2\sigma_f=1/\Tstar$) versus $\bias$ for crystal configuration configI and (d) configII. $\theta$ is the misalignment angle between crystal $c-$axis and the external magnetic field $B_c$. (e) A sketch of the dc current flow and relative positions of the crystal in two configurations. (f) Measurement scheme to reverse the dephasing from $\bias$. The $\bias$ pulse of same polarity is applied for the entirety of the sequence except near control pulses, their ringdown and near echo. (g) The echo magnitude decay measured in configI using the scheme shown in panel (f).  }
\end{figure*}
In this section, we estimate the magnetic field gradient generated by $\bias$ currents needed to tune the resonator frequency. To this end, the Hahn echo protocol is used as a probe for deducing the associated inhomogeneous broadening on the spin Larmor's frequency. We apply $\bias$ pulses for varying duration $\Delta t$ between the $\pi$ pulse and the echo [see the measurement protocol in Fig.~\ref{Sfig:dephasing}(a)] which leads to progressive dephasing of spins and reduction in echo amplitude as $\Delta t$ is increased. The decay of echo magnitude at $\bias = 2.4~$mA is plotted in Fig.~\ref{Sfig:dephasing}(b) for two crystal configurations. We model the echo decay using a two rate expression

\begin{equation}
A_e = \exp[-\left( \frac{2\tau}{T_2^0}\right)^x-\left( \frac{\Delta t}{\Tstar} \right)^2], \end{equation}
where $T_2^0=2.2~$ms and $x=1.6$ are spin-decoherence and stretch exponent deduced from measurements at $\bias=0$. From the fits shown in in Fig.~\ref{Sfig:dephasing}(b), we extract $\Tstar=200~\mu$s (12~$\mu$s) for configI (configII). The measurements are performed at various $\bias$ amplitudes and the deduced inhomogeneous broadening $2\sigma_f=1/\Tstar$ is plotted in Fig.~\ref{Sfig:dephasing}(c,d) for two configurations, showing an approximate linear dependence. 

We now discuss a simple model to reproduce the measurements. Since the exact position of spins is difficult to know in absence of input power calibration, our model is crude and only aims at a qualitative understanding. Firstly, the magnetic field $\Delta B_I$ from $\bias$ at a certain `guessed spin-position' is obtained by finite element COMSOL simulations, for which we assume that DC current flows at the edge of the ground plane  [see dashed lines in Fig.~\ref{Sfig:dephasing}(e)]. This field $\Delta B_I$ is then translated to change in transition frequency by diagonalizing the $^{167}$Er spin Hamiltonian, already in an external magnetic field $B_0=348.7~$mT. Next, we assume that all spins within the guessed detection volume contribute equally to the echo amplitude.

For configII, the $\bias$ current path directly below the crystal forms the dominant source of $\Delta B_I$. We find that the standard deviation of frequency shifts $2\sigma_f$ for spins located at a height $2.7 \pm 1 ~\mu$m is able to describe the measurements [solid line in Fig.~\ref{Sfig:dephasing}(d)]. Here we have included the misalignment angle between crystal $c-$axis and external field of 3.3~degrees deduced from continuous wave spectroscopy. In contrast, a $\Delta B_I$ strictly in the $ab$ plane of the crystal would produce quadratic change in $\sigma_f$ versus $\bias$ (dashed line) which is not consistent with measurements.

For configI, the horizontal $\bias$ current path in the lower ground plane is the main source of $\delta B_I$ [Fig.~\ref{Sfig:dephasing}(e)], which varies across the crystal-coverage (distance of $35-285~\mu$m). We find that in this case spins located at a height $16~\mu$m are needed to describe the measured $2\sigma_f$. The discrepancy in height for two configurations (by a factor of 6) is understandable due to crudeness of the model, two separate mounting of the crystal, and different powers involved in the measurement leading to selection of different spin-packets~\cite{ranjan_pulsed_2020}. 

We have shown in the main text that the dephasing of $\bias$ can be reversed when applying it on either side of the refocusing pulse for equal duration. Similar to Fig.~5, we check this reversal with long duration of $\bias$ pulses [Fig.~\ref{Sfig:dephasing}(f)] in configI. As shown in Fig.~\ref{Sfig:dephasing}(f), echo decay for $\bias=1.6~$mA shows no-visible difference compared to the case with $\bias=0$ and this validates the refocusing of static inhomogeneities achieved in the Hahn echo sequence. Long detuning pulses will be important to avoid Purcell limited spontaneous emission of spins.

%################# Retrieval efficiency
\section{Memory efficiency}
\begin{figure}[t!]
  \includegraphics[width=\columnwidth]{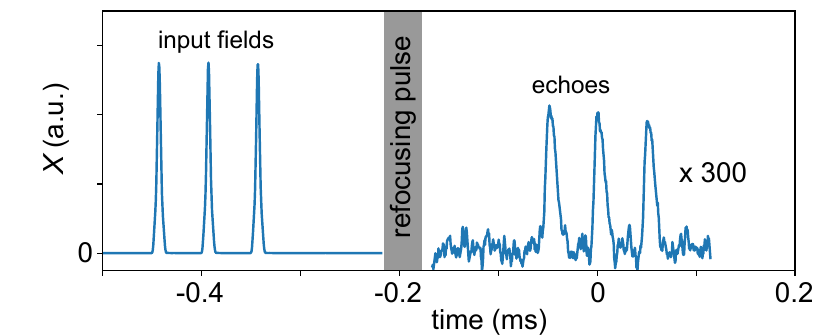}
  \caption {\label{Sfig:efficiency}
    Comparison of amplitudes of three Gaussian input fields (FWHM = $4~\mu$s) retrieved after 0.4~ms. Signals have been moved to single quadrature. Echoes are averaged over 1000 times at a repetition rate of 2s. }
\end{figure}

To estimate the memory efficiency, we compare the amplitudes of input fields with retrieved echoes [see Fig.~\ref{Sfig:efficiency}]. The input fields (Gaussian with FWHM of $4~\mu$s and relative Rabi angle of $\pi/20$) are obtained by simply measuring off resonance in the same setup such that gain calibration is not needed. Taking the ratio of integrated fields and including decoherence at 0.4~ms, we find an average field retrieval efficiency of $7 \times 10^{-3}$. Note that echoes are roughly two times wider than input fields because of filtering from comparable bandwidths of the resonator with input fields. The theoretical limits for memory efficiency used in the main text are derived in details in Ref.~\onlinecite{afzelius_proposal_2013} for the case of input-field bandwidth $\ll \kappa$.

%##########    Bibliography   ######################
%\bibliography{QuantumMemory}

%\end{document}

\end{document}